\documentstyle[mnextra]{mn}
\input{epsf}
\seceqnum           
\def\etal{{\rm et al.} }
\def\kms{{\rm km \, s^{-1}}}
\def\Mpc{{\it h}^{-1}\, {\rm Mpc}}

\def\ie {{\rm i.e. }}

\def\lsim{\mathrel{\hbox{\rlap{\hbox{\lower4pt\hbox{$\sim$}}}\hbox{$<$}}}}
\def\gsim{\mathrel{\hbox{\rlap{\hbox{\lower4pt\hbox{$\sim$}}}\hbox{$>$}}}}
\def\tcdm{$\tau$CDM}
\def\lcdm{$\Lambda$CDM}

\begin{document}
\title[Galaxy specific angular momenta]
{The Cosmological Dependence of Galactic Specific Angular Momenta}
\author[Eke, Efstathiou \& Wright]{Vincent Eke,  George Efstathiou \& Lisa 
Wright\\
Institute of Astronomy, Madingley Road, Cambridge CB3 OHA.}

\maketitle

\begin{abstract}
Hydrodynamical simulations of galaxy formation in spatially flat Cold
Dark Matter (CDM) cosmologies with and without a cosmological constant
($\Lambda$) are described.  A simple star formation algorithm is
employed and radiative cooling is allowed only after redshift $z=1$ so
that enough hot gas is available to form large, rapidly rotating
stellar discs if angular momentum is approximately conserved during
collapse. The specific angular momenta of the final galaxies are found
to be sensitive to the assumed background cosmology. This dependence
arises from the different angular momenta contained in the haloes at
the epoch when the gas begins to collapse and the inhomogeneity of the
subsequent halo evolution. In the $\Lambda$-dominated cosmology, the
ratio of stellar specific angular momentum to that of the dark matter
halo (measured at the virial radius) has a median value of $\sim 0.24$
at $z=0$.  The corresponding quantity for the $\Lambda=0$ cosmology is
over $3$ times lower. It is concluded that the observed frequency and
angular momenta of disc galaxies pose significant problems for
spatially flat CDM models with $\Lambda=0$ but may be consistent with
a $\Lambda$-dominated CDM universe.
\end{abstract}

\section{Introduction}\label{sec:intro}

Smoothed Particle Hydrodynamical (SPH) simulations of galaxy formation
in Cold Dark Matter (CDM) dominated universes have repeatedly failed
to create Milky Way-type extended discs (Navarro \& Benz 1991; Katz
1992; Navarro, Frenk \& White 1995; Steinmetz \& Muller 1995; Navarro
\& Steinmetz 1997; Weil, Eke \& Efstathiou 1998, hereafter WEE98). 
Gas is found to cool
very effectively at high redshifts into the centres of haloes before
gravitational torques from surrounding perturbations can supply it
with sufficient angular momentum. Furthermore the lumpy nature of the
subsequent halo evolution leads to an outward transfer of angular
momentum, compounding the problem. Navarro \& Steinmetz (1997) showed
that including a photoionizing background to suppress the
early cooling of gas, worsened the problem by preferentially
decreasing the amount of higher angular momentum gas accreted at late
epochs. 


Almost all analytic and semi-analytic models of galaxy formation
assume that angular momentum is conserved during disc formation
(Mestel 1963; Fall \& Efstathiou 1980; Gunn 1982; van der Kruit 1987;
Cole \etal 1994; Dalcanton, Spergel \& Summers 1997; Mo, Mao \& White
1998), in marked contrast to what is found in numerical
simulations. It is interesting to note that Cole \etal (1994) actually
find that their predicted Tully-Fisher relation has galaxies spinning
$60$ per cent too rapidly at fixed luminosity, or alternatively being
underluminous at fixed circular velocity. Evidently, the degree to
which angular momentum is conserved during galaxy formation is of
vital importance in developing realistic theoretical models for 
comparison with observations.

Given that high angular momentum gas is required to produce a rapidly
spinning large stellar disc, and recognising that the collapse of gas
to form such a disc will, if anything, transfer angular momentum to
the halo, it is apparent that a significant reservoir of hot gas must
be maintained in the galactic halo at least until the epoch at which
the halo specific angular momentum has grown to match that of real
galaxies. As discussed above, numerical simulations have generally
failed to do this. Navarro \& Benz (1991),  WEE98 and others,
have suggested that the inclusion of a more effective
feedback mechanism, in particular, energy injection from supernovae
(see {\it e.g.} Larson 1974, Dekel and Silk 1986) is the most likely
solution to this problem. To illustrate the importance of feedback, WEE98
performed several SPH simulations of haloes selected from CDM initial
conditions in an $\Omega=1$, $\Lambda = 0$, universe with radiative
cooling suppressed until a redshift $z=1$. They did indeed find that
suppressing the collapse of gas until late epochs had a dramatic
effect on the specific angular momentum of the final
galaxies. However, only two out of five carefully selected haloes
produced objects with comparable angular momenta to those of real disc
galaxies. This suggests that even with an extreme feedback
prescription, there may be a problem explaining the frequency and
angular momenta of disc galaxies in a critical density CDM universe
with $\Lambda=0$.

This problem is investigated in more detail in this letter and the
sensitivity of the specific angular momenta of the final objects to
the assumed background cosmology is tested. The same
simplified feedback prescription of WEE98 is adopted, {\it i.e.}
radiative cooling is suppressed until $z=1$, and applied to a total
of $60$ SPH simulations for two spatially flat CDM universes, one with
$\Omega_m=1$ and $\Lambda=0$ and the other with $\Omega_m = 0.3$ and
$\Omega_\Lambda = 0.7$. Section~\ref{sec:sims} contains details of the
two cosmological models, the initial conditions of the SPH simulations
and the simulations themselves.  The results are presented in
Section~\ref{sec:res} and discussed in Section~\ref{sec:conc}.

\section{Simulation details}\label{sec:sims}

\subsection{Model parameters}\label{ssec:models}

The two spatially flat CDM cosmological models that are investigated
have scale invariant initial fluctuations and a
post-recombination power spectrum given by the parametrisation of
Efstathiou, Bond \& White (1992) with a shape parameter
$\Gamma=0.2$. One model, referred to as \tcdm ~(following the
nomenclature of Thomas \etal 1998) has a matter density parameter $\Omega_m
= 1$ and zero cosmological constant.  The other model, 
\lcdm ~has $\Omega_m = 0.3$ and a cosmological constant
contributing $\Omega_\Lambda = 0.7$ to the density parameter; the
parameters for this model are close to those favoured by anisotropies
in the cosmic microwave background radiation and the distances of Type
 Ia supernovae (see {\it e.g.} Efstathiou \etal 1999). The amplitude of
the mass fluctuations has been normalised to reproduce the present day
abundance of galaxy clusters, thus the linear theory {\it rms} mass
fluctuations in spheres of radius $8\Mpc$ are $\sigma_8=0.52$ and
$0.90$ for the \tcdm ~and \lcdm ~models respectively (Eke \etal 1996).
In both models the Hubble constant is set to $h=0.65$\footnote{$h$ is
defined such that $H_{0}=100h~ \kms{\rm Mpc}^{-1}$.}, giving an age
for the \tcdm ~universe of $10.0$~Gyr and $14.5$~Gyr for \lcdm.
The baryonic contribution to the critical density is set to
$\Omega_{\rm b}=0.06$ in both models, consistent with the predictions
of primordial nucleosynthesis and the deuterium abundance measurements
reported by  Burles \& Tytler (1998). 

\subsection{Dark matter simulations}\label{ssec:ap3m}

To create initial conditions for the SPH simulations, a dark matter
only calculation was performed for each cosmology using the AP$^3$M code
of Couchman (1991). For each cosmology, a $32.5$ $\Mpc$~cube
containing $128^3$ particles was evolved from $z=24$ to the present
employing $2000$ timesteps of equal size in expansion factor. The
effective Plummer gravitational force softening was fixed in comoving
coordinates at $7$ kpc.  Identical random phases were used in both
simulations so that the same haloes in both cosmologies could be
simulated at higher resolution with the SPH code.

\subsection{Halo selection}\label{ssec:halo}

The spherical overdensity group-finding algorithm (Lacey \& Cole 1994)
was applied to the final outputs of the dark matter simulations to
locate virialised haloes, with the virial radius defined by the
spherical collapse model. A subset of these were chosen for resimulation
with the SPH code. To qualify for resimulation, a halo had to be at least $1
\Mpc$ from any other containing at least $100$ particles at
$z=0$. This constraint was applied so that the effort in the
resimulation was concentrated on the central object rather than a more
massive companion. It effectively 
biases against haloes in dense environments and can be loosely thought of
as selecting a sample of  field galaxies.  Twenty 
haloes that could clearly be identified as the same object in both
\tcdm ~and \lcdm ~simulations with  circular velocities in the range
$170<v_{\rm c}/~\kms<250$ in the \tcdm ~run were selected for resimulation
(referred to as `common' haloes hereafter).
As the corresponding \lcdm ~haloes had systematically lower circular
velocities (by about $30$ per cent), 
$5$ additional larger \lcdm ~and $15$ smaller \tcdm
~haloes were also chosen to increase the overlap in circular
velocity between the two models. WEE98 adopted similar
algorithms to select haloes, but also imposed an additional constraint
that the haloes should not have merged with a comparable mass system
between $z=1$ and $z=0$. This criterion was imposed to bias against
haloes that suffered a major merger at late times and hence to
select haloes more favourable to the formation of disc systems. No such
criterion was applied in generating initial conditions for
the simulations described in this paper.

\subsection{SPH Simulations}\label{ssec:resim}

The procedure for creating high resolution initial conditions for
resimulation is as described by WEE98. Briefly, this involved tracing
back to the initial redshift all dark matter particles within
$400$~kpc of the selected halo centres at $z=0$. Extra particles were
placed in a `high resolution' cube containing the region of interest
(and including the short wavelength fluctuations associated with this
improved resolution) and more massive particles were added to 
sample the distant density field.  $34^3$ dark matter and $34^3$ gas
particles (initially at identical positions to the dark particles)
were used in the central cube, while the outer regions were
represented by $\sim 5000$ particles with radially increasing
masses. The sizes of the high resolution cubes were typically about
$3.2\Mpc$ and $4.0\Mpc$ for \tcdm ~and \lcdm ~simulations, yielding
gas particle masses of $\sim 2\times10^7{\rm M_\odot}$ and
$4\times10^7{\rm M_\odot}$ respectively. Higher resolution runs with
$2\times 43^3$ particles were also performed for some of the \tcdm
~haloes that produced inadequately resolved stellar discs.

The evolution of the simulation was performed using the GRAPESPH code
outlined in WEE98 and $5$ GRAPE$-3A$ boards (Sugimoto \etal 1990)
connected to a Sun Ultra$-2$ workstation. The Plummer gravitational
force softening for gas and star particles was $0.8$~kpc and the dark
matter had softenings of $4.1$~kpc and $2.7$~kpc for \tcdm ~and \lcdm
~respectively.  Up to $40000$ timesteps for \tcdm ~and $60000$ for
\lcdm ~runs were used to evolve the particles from $z=24$ to $z=0$.
Typical run times were $2$ days for each \tcdm ~simulation and $3-4$
days for a \lcdm ~simulation.
Radiative cooling was switched on at $z=1$ in all cases to
model feedback crudely, as described by WEE98.  Each gas particle that
remained in a collapsing region with $\rho>7\times10^{-23}$ kg
m$^{-3}$ (see Navarro \& White 1993, WEE98) for a local dynamical time was
converted to a star particle.

\section{Results}\label{sec:res}

\begin{figure}
\centerline{\epsfxsize=9cm \epsfbox{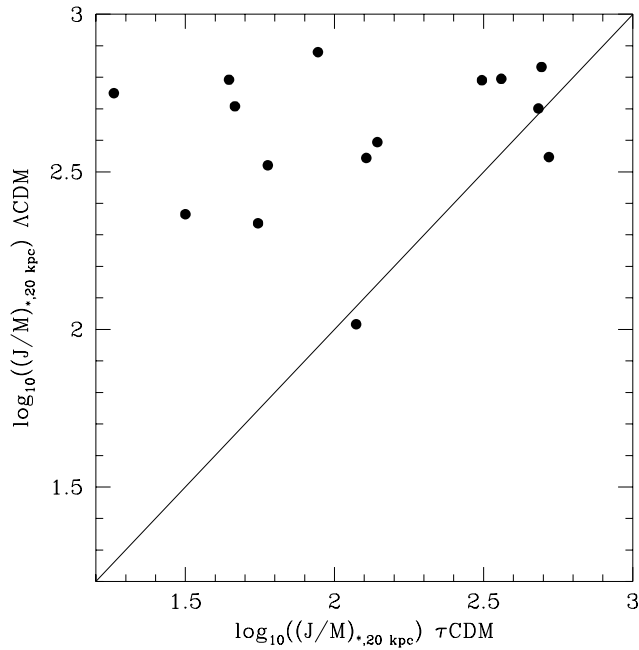}}
\caption{The correlation of the stellar specific 
angular momenta for the \tcdm ~and \lcdm ~SPH simulations of `common'
haloes ({\it i.e.} haloes clearly identified as the same object in 
both cosmologies, see Section~\ref{ssec:halo}). 
The stellar specific angular momentum
is computed for stars within $20$kpc of the stellar centre.
Results are shown only for systems that contain more than $1000$
stars.
}
\label{fig:dircomp}
\end{figure}

The same number of particles is used in the high resolution regions
even though these differ in volume from run-to-run. The numerical
resolution of the SPH simulations is therefore variable and correlated
with the mass of the halo.  Navarro \& Steinmetz (1997) discussed in
detail the outward transport of angular momentum when the SPH
artificial viscosity acts on a poorly resolved gaseous disc. As a
result of the star formation algorithm adopted here, the gaseous discs
do not monotonically increase their masses, and thus the effect of
this viscous transport is probably increasing as the simulations
approach $z=0$ and the accretion rate diminishes. However, by
analysing the conservation of angular momentum as a function of the
number of star particles within $20$ kpc of the halo centre, it was found
that only systems with less than $\sim 1000$ stars showed any
significant trend of increasing angular momentum with increasing mass
resolution. Consequently, results will only be given for simulations
that had at least $1000$ stars in the central object at $z=0$.

\begin{figure}
\centerline{\epsfxsize=9cm \epsfbox{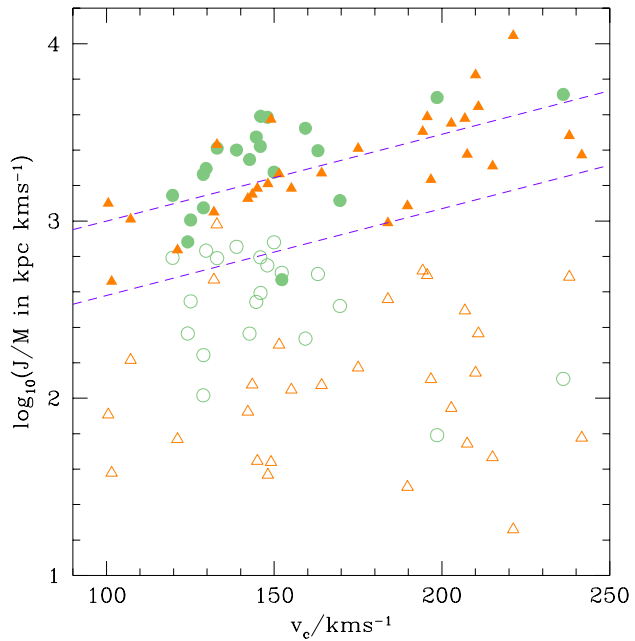}}
\caption{Specific angular momenta of the dark matter haloes,
determined at the virial radius, and the stellar components, at $20$
kpc. \tcdm ~haloes and stellar components are represented by filled
and open triangles, while the corresponding \lcdm ~variables are shown
with filled and open circles. The dotted lines delineate the
approximate $90$ percentile range of specific angular momenta of
observed disc systems derived from the sample analysed by WEE98,
converted to a Hubble constant of $h=0.65$ and assuming that the discs
are perfectly cold and have flat rotation curves.}
\label{fig:jom}
\end{figure}

Figure~\ref{fig:dircomp} shows an object by object comparison of the
stellar specific angular momenta for the
common haloes in the two sets of simulations. The stellar angular
momentum is measured for the central galaxy-like object within $20$
kpc of its centre. There is essentially no correlation
between the two quantities. However, the galaxies in the \lcdm ~simulations 
tend to have significantly 
more angular momentum than their \tcdm ~counterparts. The
reasons for this difference will be discussed in more detail below.

\begin{figure}
\centerline{\epsfxsize=9cm \epsfbox{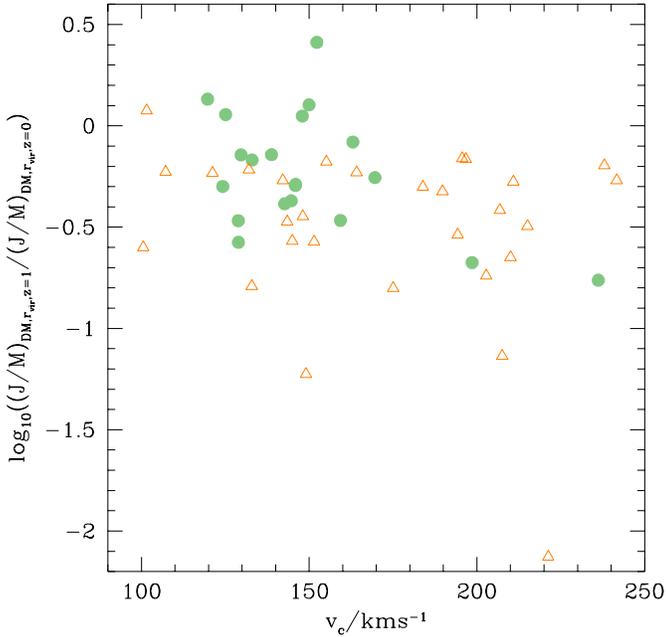}}
\caption{The ratio of $z=1$ to $z=0$ halo specific angular 
momenta for \tcdm ~(open triangles) and \lcdm ~(filled circles) simulations
plotted as a function of circular velocity.
}
\label{fig:dmjomrat}
\end{figure}

Figure~\ref{fig:jom} shows the absolute specific angular momenta of all
the dark matter haloes (\ie not just the common haloes), measured at
the virial radius, and their largest stellar occupants, as a function
of halo circular velocity. The dashed lines in Figure~\ref{fig:jom} show
the ranges of specific angular momenta of real disc
galaxies\footnote{The observed specific angular momenta are calculated
assuming flat rotation curves of amplitude $v_c$. For dark haloes with
the Navarro, Frenk and White (1996) profile, the disc rotation
velocity may exceed the halo circular speed at the virial radius by as
much as $30$--$40$ \%, depending on the concentration of the stellar
disc. This difference is neglected in this paper.} (see figure 1 of
WEE98).  While the haloes in both cosmologies occupy a similar locus
(for $120 < v_{\rm c}/\kms < 180$, the \lcdm ~haloes are only $\sim
50$ per cent higher than those for \tcdm), the \tcdm ~stellar objects
are at systematically much lower values than those from the \lcdm
~simulations. For both sets of simulations the fraction of halo
specific angular momentum retained in the stellar objects decreases
with increasing circular speed.  Thus in comparing the two
cosmologies, attention will be restricted to haloes with circular
speeds in the same range, $120 < v_{\rm c}/\kms < 180$.  For the $12$
\tcdm ~and $17$ \lcdm ~haloes that this range includes, the median
ratios of stellar to halo specific angular momenta at $z=0$ are $0.07$
and $0.24$ respectively.

One reason for this difference is that although haloes in the \tcdm
~and \lcdm ~models with the same circular speed have similar angular
momenta at $z=0$, the angular momentum growth depends on the
background cosmology. As a consequence of the continual evolution of
structure in the \tcdm ~model, haloes acquire relatively more angular
momentum since $z=1$ than haloes in the \lcdm ~model.  This is
illustrated in Figure~\ref{fig:dmjomrat}.  Since typically half of the
stars are formed by $z=0.4$ in the \tcdm ~simulations and by $z=0.6$
for \lcdm, it is more appropriate to compare final stellar angular
momenta with those of the parent haloes at $z=1$ when disc formation
begins.  For haloes with circular speeds in the range $120 <
v_{\rm c}/\kms < 180$ the median ratios of stellar to halo
specific angular momenta at $z=1$, are $0.17$ and $0.32$ for \tcdm
~and \lcdm ~respectively.  Thus a major reason for the differences in
final stellar angular momenta in the two cosmologies is that \tcdm
~haloes have lower angular momenta at $z\sim 1$ compared with \lcdm
~haloes of the same circular speed.

\begin{figure}
\centerline{\epsfxsize=9cm \epsfbox{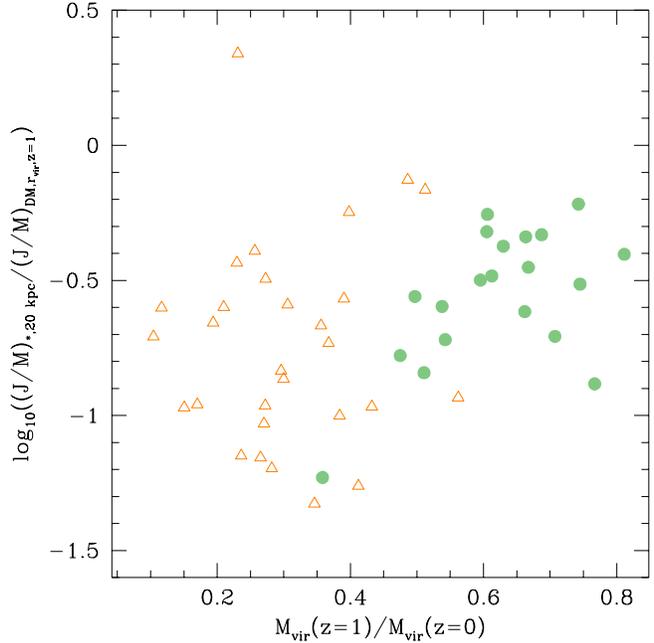}}
\caption{The correlation between the ratio of halo 
virial mass at $z=1$ to $z=0$
and the conserved specific angular momentum.
\tcdm ~results are shown with open triangles and \lcdm ~with filled circles.
}
\label{fig:mvirrat}
\end{figure}

The other main cause of the difference in stellar angular momenta in
the two cosmologies stems from the more inhomogeneous evolution at $z
< 1$ suffered by \tcdm ~haloes.  As a consequence of the higher
frequency of merger events in the \tcdm ~cosmology, the accretion at
late times of  high angular momentum gas from the outer parts of
the halo is disrupted. Either the gas is gravitationally shocked and
remains extended, or its angular momentum is transported to larger
radii by torques from the anisotropic halo. Figure~\ref{fig:mvirrat}
shows how the fraction of conserved specific angular momentum varies
with the halo growth, parametrised by the ratio of halo virial masses
at $z=1$ to $z=0$. It is clear that the \lcdm ~haloes accrete
significantly less mass than those evolving in the \tcdm
~model. Galaxies forming in the \lcdm ~cosmology experience fewer
merger events and conserve a greater fraction of their angular momentum.

\section{Conclusions}\label{sec:conc}

\begin{figure}
\centerline{\epsfxsize=9cm \epsfbox{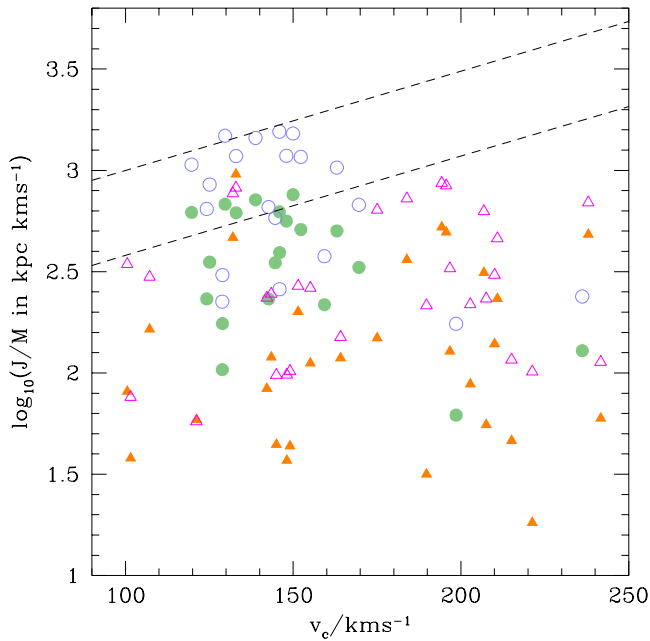}}
\caption{The stellar specific angular momenta for \tcdm ~and \lcdm ~galaxies
are shown with filled triangles and circles, as a function of circular 
velocity. The corresponding open symbols represent the results obtained when
stars within a central sphere of $3$ kpc are excluded from the calculation.
Dashed lines show the range of specific angular momenta of real disc
galaxies as plotted in Figure~\ref{fig:jom}.}
\label{fig:hole}
\end{figure}

The aim of this work has been to investigate how the background
cosmological model affects the efficiency with which angular momentum
is conserved during galaxy formation.  The large set of simulations
described here shows that the specific angular momenta of galaxies with
halo circular speeds in the range $120$--$180\; \kms$ forming in a
\lcdm ~universe are typically $3-4$ times higher than those in a \tcdm
~universe. This large difference arises from the more turbulent merger
histories of the \tcdm ~haloes and from their lower specific angular
momenta.

The median specific angular momentum for observed disc galaxies (see
the dashed lines in Figure~\ref{fig:jom} and figure 1 of WEE98) is,
nevertheless,  about $2.5$ times as large as that of the \lcdm
~galaxies simulated here, and an order of magnitude above the \tcdm
~galaxies.  This discrepancy for the \tcdm ~model is extremely
large, confirming the suspicion of WEE98 that the observed frequency
of disc galaxies is difficult to reproduce in a spatially flat CDM
model with $\Lambda=0$.  Allowing cooling at higher redshift would
exacerbate this problem (see WEE98), and it seems unlikely that a
more realistic feedback process could resolve this
discrepancy.  

In both cosmologies there is a strong trend for galaxies with high
circular speeds to lose more angular momentum, suggesting that it is
difficult to form large disc galaxies within haloes with circular
speeds of $\gsim 200 \kms$ at the virial radius. This is qualitatively
consistent with the observed lack of spiral discs with high circular
speeds.

Most of the simulated galaxies in the \lcdm ~cosmology lie below the
observed range of specific angular momenta for real disc systems,
though the magnitude of the discrepancy is much lower than in the
\tcdm ~cosmology. This may not be an insurmountable problem for the
\lcdm ~model. The observational range plotted in Figure~\ref{fig:jom}
is computed from the specific angular momenta of only the disc
components, whereas the specific angular momenta of the simulated
galaxies have been calculated using the entire stellar system. In a
more realistic scenario, the low angular momentum gas might be
expected to give rise to a bulge-type stellar component in addition to
providing the feedback energy to maintain the reservoir of hot,
extended and high angular momentum gas at large redshift.
Figure~\ref{fig:hole} shows that removing all stars within a $3$ kpc
sphere of the centres of the final stellar objects can bring the \lcdm
~galaxies with circular speeds $\lsim 180 \kms$ into agreement with
the observations. The choice of $3$ kpc is arbitrary and has been
adopted merely to illustrate the importance of correctly
distinguishing between an inner bulge and an extended disc before
comparing simulations with observations.  To do this more accurately
will require a more realistic feedback prescription and much larger
numerical simulations in which bulge and disc components can be
distinguished kinematically.  Until such simulations are performed, it
is unclear whether the \lcdm ~model can account for the observed
frequency of disc galaxies. However, galaxies forming in a \tcdm-like
universe experience a severe `angular momentum catastrophe' and this
seems to be a fundamental problem for such a model.

\section*{ACKNOWLEDGMENTS}

VE, GE and LW acknowledge the support of a PPARC postdoctoral
fellowship, senior research fellowship and research studentship
respectively. We thank the Institute of Astronomy for providing funds
towards the purchase of the GRAPE boards.

\end{document}